\documentclass{article}
\usepackage{hiph-art}
\usepackage{epsfig}
\usepackage{graphicx}
\usepackage{amsmath}
\usepackage{amssymb}
\volnumber{19} \issuenumber{1} \edyear{2004}                             
\frompage{000} \topage{000}                                              
\recrevdate{1 January 2004}                                              
\def\rightmark{Ellipsoidal flows in relativistic hydrodynamics of finite systems}
\def\leftmark{Yu.M. Sinyukov et al.}

\title{Ellipsoidal flows in relativistic hydrodynamics of finite systems}
\authors{
{Yu.M.  Sinyukov$^{1}$ and Iu.A.Karpenko$^{1,2}$ %
\index{One, A.} 
\index{Two, A.} 
}\\[2.812mm]
{\normalsize \hspace*{-8pt}$^1$ Bogolyubov Institute for Theoretical
Physics, Kiev 03143, Metrologichna 14b, Ukraine\\[0.2ex]
\hspace*{-8pt}$^2$ National Taras Shevchenko University of Kyiv,
Kiev 01033, Volodymyrska 64, Ukraine.
}}

\abstract{A new class of 3D anisotropic analytic solutions of
relativistic hydrodynamics with constant pressure is found. We
analyse, in particular, solutions corresponding to ellipsoidally
symmetric expansion of finite systems into vacuum. They can be
utilized for relativistic description of the system evolution in
thermodynamic region near the softest point and at the final stage
of the hydrodynamic expansion in A+A collisions. The solutions can
be used also for testing of numerical hydrodynamic codes solving
relativistic hydrodynamic equations for anisotropic expansion of
finite systems.}



\makeindex
\begin{document}

\maketitle

\section{Introduction}
The solutions of relativistic hydrodynamics are used typically in
cosmology, astrophysics and for an analysis of the processes of
ultra-relativistic heavy ion collisions. It is remarkable fact
that in these so different fields the Hubble-like solutions play
the important role. As for A+A collisions the Hubble-like models
becomes to be popular for a description of the experimental data
 only lately \cite{flor} since it was found in
RHIC experiments that rapidity distribution is not flat even at
these top energies, as it was expected (see, e.g., rewiev in
\cite{Busza}). Thus, one can suppose that the initial state could
be more close to spherically symmetric one rather then to
boost-invariant in longitudinal direction \cite{Hwa,Bjorken}.

The spherically symmetric hydrodynamic solutions with the Hubble
velocity distribution, $v=r/t$, has been considered for the first
time in Ref. \cite{Chiu}. The equation of state (EoS) was chosen
as ultrarelativistic one: $p=c_{0}^{2}\varepsilon$. Some
generalization of these results was proposed in a case of the
Hubble flow for EoS of massive gas with conserved particle number
in Ref.\cite{Csorgo}.

It is naturally, however, that, unlike to the Hubble type flows,
the velocity gradients in thermal matter formed in A+A collisions
should be different in different directions since there is an
initial asymmetry between longitudinal and transverse directions
in central collisions and, in addition, between in-plane and
off-plane transverse ones in non-central collisions. The another
important remark is that the Hubble-like  hydrodynamic solutions
is related to infinite systems while the matter in A+A collisions
is occupied essentially finite region.

In this letter we search for analytical solutions of the
hydrodynamic equations describing 3D asymmetric relativistic
expansions of finite systems.

\section{General analysis}
\ \ \ \ Let us start from the equations of relativistic
hydrodynamics:
\begin{equation}\label{gener}
    \partial_{\nu}T^{\mu\nu}=0,
\end{equation}
where the energy-momentum tensor corresponds to a perfect fluid:
\begin{equation}\label{ideal_fluid}
    T^{\mu\nu}=(\varepsilon+p)u^{\mu}u^{\nu}-p\cdot g^{\mu\nu}
\end{equation}
We can attempt to find a particular class of solutions and therefore
have to make some simplifications of (\ref{gener}).

Let us put pressure to be constant,
\begin{equation}\label{p=const}
   p=p_0=const,
\end{equation}
in other words, we are searching for the solution near the softest
point where the velocity of sound $c_{s}=0$.

Then we find from Eq.(\ref{gener})
\begin{equation}
    (\epsilon+p)\partial_{\nu}(u^{\mu}u^{\nu})+u^{\mu}u^{\nu}\partial_{\nu}\epsilon=0
\end{equation}
Contracting the latter equation with $u_{\mu}$ we obtain:
\begin{equation}\label{second eq}
    (\varepsilon+p)\partial_{\nu}u^{\nu}+u^{\nu}\partial_{\nu}\varepsilon=0.
\end{equation}
and, thus, the remaining equation to satisfy is
\begin{equation}\label{first eq}
   u^{\nu}\partial_{\nu}u^{\mu}=0,
\end{equation}
which means that flow is accelerationless in the rest systems of
each fluid element; this property holds for the known Bjorken
(boost-invariant) and  Hubble flows.

Finally, the relativistic hydrodynamics at the softest point is
described by the equations (\ref{second eq}) and (\ref{first eq})
for the hydrodynamic velocities $u^{\mu}$, and energy density. A
serious problem is, however, to find non-trivial solutions for the
field $u^{\mu}(x)$ of hydrodynamic 4-velocities.

\section{Relativistic ellipsoidal solutions}
\ \ \ \ First the solution at $p=const$ was obtained in
\cite{Biro} as physically corresponding to a thermodynamic state
of the system in the softest point with the velocity of sound
$c_{s}^{2}=0$. Such a state could be associated with the first
order phase transition. In A+A collisions it corresponds,
probably, to transition between hadron and quark-gluon matter at
SPS energies. The solution proposed in \cite{Biro}  has the
cylindrical symmetry in the transverse plane and the longitudinal
boost invariance:
\begin{equation}\label{vb}
    u_{\mu}=\gamma(\frac{t}{\tau},v\frac{x}{r},v\frac{y}{r},\frac{z}{\tau}),
\end{equation}
where $\tau=\sqrt{t^{2}-z^{2}}$, $\gamma=(1-v^{2})^{-1/2}$, $r$ is
transverse radius, $r=\sqrt{x^{2}+y^{2}}$, and
\begin{equation}\label{vv}
    v=\frac{\alpha}{1+\alpha\tau}r
\end{equation}
describes axially symmetric transverse flow.

The above solution has, however, a limited region of applicability
since the boost invariance is not expected at SPS energies and can
be used only in a small mid-rapidity interval \cite{AkkMunSin}, it
is not reached even at RHIC energies \cite{Busza}. But most
important is that in non-central collisions there is no axial
symmetry and, therefore, one needs in transversely asymmetric
solutions to describe the elliptic flows in these collisions,
e.g., $v_2$ coefficients. Now we propose a new class of analytic
solutions of the relativistic hydrodynamics for 3D asymmetric
flows.

Firstly we construct the ansatz for normalized 4-velocity:
\begin{equation}
u^{\mu}=\{\frac{t}{\sqrt{t^{2}-\sum a_{i}^{2}(t)x_{i}^{2}}},\frac
{a_{k}(t)x_{k}}{\sqrt{t^{2}-\sum a_{i}^{2}(t)x_{i}^{2}}}\}\label{elliptic sol}%
\end{equation}
where the Latin indexes denote spatial coordinates, which are
functions of time only. In this case
a set of nonequal $a_{i}$ induces 3D elliptic flow with velocities $v_{i}%
=a_{i}(t)x_{i}/t$: at any time $t$ the absolute value of velocity is
constant,
$\mathbf{v}^{2}=const$, at the ellipsoidal surface $\sum a_{i}^{2}x_{i}%
^{2}=const$.

The condition (\ref{first eq}) of accelerationless thus reduces to
the ordinary differential equation (ODE) for the functions
$a_{i}(t)$:
\begin{equation}
\frac{da_{i}}{dt}=\frac{a_{i}-a_{i}^{2}}{t}, \label{ak ode}%
\end{equation}
the general solution of which is:
\begin{equation}
a_{i}(t)=\frac{t}{t+T_{i}}, \label{ak}%
\end{equation}
where $T_{i}$ is some set of 3 parameters (integration constants)
having the dimension of time. The different values $T_{1}$, $T_{2}$
and $T_{3}$ results in anisotropic 3D expansion with the ellipsoidal
flows.

The solution of Eq. (\ref{second eq}) for energy density
$\varepsilon$ is found as follows. Taking into account that
$\partial_{\mu}u^{\mu}=\sum a_{i}/\widetilde{\tau}$, where
\begin{equation}\label{tilde}
    \widetilde{\tau }=\sqrt{t^{2}-\sum a_{i}^{2}x_{i}^{2}},
\end{equation}
one can get
\begin{equation}\label{eq-enthalpy}
(\varepsilon+p_{0})\sum_{i}a_{i}(t)+t\partial_{t}\varepsilon+\sum_{i}a_{i}(t)x^{i}%
\partial_{i}\varepsilon=0.
\end{equation}
General solution of the equation is
\begin{equation}\label{general}
  \varepsilon+p_{0}= \frac{F_{\varepsilon}(\frac {x_{1}}{t+
T_{1}},\frac {x_{2}}{t+T_{2}},\frac {x_{3}}{t+T_{3}})} {(
t+T_{1})(t+T_{2})(t+T_{3})}
\end{equation}
where $F_{\varepsilon}$ is an arbitrary function of its variables.
At fixed $T_{i}$ that define the velocity profile, function
$F_{\varepsilon}$ is completely determined by the initial conditions
for the enthalpy density.

If some value, associated with a quantum number or with particle
number in a case of chemically frozen evolution is conserved
\cite{AkkMunSin}, then one should add the corresponding equation
to the basic ones. Such an equation has the standard form
\cite{Landau2}:
\begin{equation}\label{n_eq}
n\partial_{\nu}u^{\nu}+u^{\mu}\partial_{\mu}n=0
\end{equation}
where $n$ is associated with density of the correspondent
conserved value, e.g., with the baryon or particle density. A
general structure of this equation is similar to Eq.
(\ref{eq-enthalpy}) and, therefore, the solution:
\begin{equation}\label{general1}
n= \frac{F_{n}(\frac {x_{1}}{t+ T_{1}},\frac {x_{2}}{t+T_{2}},\frac
{x_{3}}{t+T_{3}})} {( t+T_{1})(t+T_{2})(t+T_{3})},
\end{equation}
which looks like as (\ref{general}) and the function $F_{n}$ is
also an arbitrary function of its arguments and can be fixed by
the initial conditions for (particle) density $n$. The
corresponding chemical potential is not zero and describes the
deviation from chemical equilibrium in relativistic systems. The
thermodynamic identities lead to the following expression for the
temperature:
\begin{equation}\label{T}
    T(t,\textbf{x})=F_{T}(\frac {x_{1}}{t+
T_{1}},\frac {x_{2}}{t+T_{2}},\frac {x_{3}}{t+T_{3}})
\end{equation}
where $F_{T}$ is some function of its arguments that is defined by
the initial conditions for $\varepsilon$ and $n$ as well as by EoS
$\varepsilon=\varepsilon(n,T)$. If the initial enthalpy density
profile is proportional to the particle density profile,
$F_{n}(\frac {x_{1}}{T_{1}},\frac {x_{2}}{T_{2}},\frac
{x_{3}}{T_{3}})\sim F_{\varepsilon}(\frac {x_{1}}{T_{1}},\frac
{x_{2}}{T_{2}},\frac {x_{3}}{T_{3}})$, then $T=const$ and so
$\mu=const$.

In a case of chemically equilibrium expansion of the
ultrarelativistic gas when the particle number is uncertain and is
defined by the conditions and parameters of the thermodynamic
equilibrium, e.g., by the temperature $T$, the chemical potential
$\mu\equiv 0$. Then it follows that $T=const$ for such a system
and the entropy $s=(\varepsilon(t,\textbf{x})+p_{0})/T$ where
$\varepsilon(t,x)$ is defined by (\ref{general}).

\section{Relativistic anisotropic expansion of finite systems}
\ \ \ \ Let us describe some important particular solutions of the
equations for relativistic ellipsoidal flows. If one defines the
initial conditions on the hypersurface of constant time, say
$t=0$, then $t$ is a natural parameter of the evolution. Such a
representation of the solutions, which are similar to the Bjorken
and Hubble ones, with velocity field $v_{i}=a_{i}x_{i}/t$ has
property of an infinite  velocity increase at
$x\rightarrow\infty$. A real fluid, therefore, can occupy only the
space-time region where $|\textbf{v}|<1$, or $\tilde{\tau}^2>0$.
To guarantee the energy-momentum conservation of the system during
the evolution, all thermodynamic densities have to be zero at the
boundary of the physical region, otherwise one should consider the
boundary as the massive shell \cite{sin}. Hence in the standard
hydrodynamic approach the enthalpy and particle density must be
zero at the surface defined by $|\textbf{v}(t,\textbf{x})|=1$ at
any time $t$. One of a simple form of such a solution (for the
case of particle number conservation) can be obtained from
(\ref{general}),(\ref{general1}) choosing
$F_{\epsilon,n}\sim\exp\left(-b_{\epsilon}^{2}\frac{t^2}{\widetilde{\tau
}^2}\right)$:
\begin{eqnarray}
  \varepsilon+p_{0}=\frac{C_{\varepsilon}}{\prod_{i}(t+T_{i})}\exp\left(-b_{\epsilon}^{2}\frac{t^2}{\widetilde{\tau
}^2}\right),\label{t} \\
n=\frac{C_{n}}{\prod_{i}(t+T_{i})}\exp\left(-b_{n}^{2}\frac{t^2}{\widetilde{\tau
}^2}\right)\label{n},
\end{eqnarray}
where $\tilde{\tau}$ is defined by (\ref{tilde}), and the constants
$C_{\varepsilon}$, $C_{n}$, $b_{\epsilon}$ and $b_{n}$ are
determined by the initial conditions as described in the previous
section. As one can see, the enthalpy density tends to zero when
$|\textbf{x}|$ becomes fairly large approaching the boundary surface
defined by $|\textbf{v}(t,\textbf{x})|=1$, in the other words, when
$\tilde{\tau}\rightarrow 0$. Thus the physically inconsistent
situation when massive fluid elements move with the velocity of
light at the surface $\tilde{\tau}=0$ is avoided. Of course, in such
a solution one has to put a constant pressure to be zero, $p_{0}=0$.

On the other hand, the solution (\ref{t}), (\ref{n}) can describe
effectively finite systems at the hypersurface $\tilde{\tau}=const$.
To show this one can substitute by the definition
$t^2=\tilde{\tau}^2+\sum a_i^2x_i^2$ in the solution, and get:
\begin{eqnarray}
  \varepsilon+p_{0}=\frac{C_{\varepsilon}\exp(-b_{\epsilon}^2)}{\prod_{i}(t+T_{i})}\exp\left(-b_{\epsilon}^2\frac{\sum a_i^2x_i^2}{\widetilde{\tau
}^2}\right),\label{tt} \\
n=\frac{C_{n}\exp(-b_n^2)}{\prod_{i}(t+T_{i})}\exp\left(-b_{n}^{2}\frac{\sum
a_i^2x_i^2}{\widetilde{\tau }^2}\right),\label{nt}
\end{eqnarray}
supposing that $x_i$ are spatial coordinates on the hypersurface.
This describes effectively finite hydrodynamic flow with spatial
radii $R_i\approx\tau/b_{\epsilon}$ at $\tau\gg T_k$.

As it follows from an analysis of the behavior of the
thermodynamic values in the previous section, the temperature is
constant if $b_{\varepsilon}=b_{n}$, otherwise one can choose the
temperature approaching zero at the system's boundary, e.g., for
EoS which is linear in temperature, the latter has the form
\begin{align}
T  & =const,\text{
\ \ \ \ \ \ \ \ \ \ \ \ \ \ \ \ \ \ \ \ \ \ \ \ \ \ \ \ \ \ \ \ \ \ \ \ }%
b_{\varepsilon}=b_{n},\nonumber\\
& \label{Tdecrease}\\
T  & \sim
e^{-(b_{\varepsilon}^{2}-b_{n}^{2})\frac{t^{2}}{\tilde{\tau}^{2}}}\rightarrow
0, \ \ \ |v(x)|\rightarrow 1, \ \ \ \ \ b_{\varepsilon}>
b_{n}\nonumber.
\end{align}

Note that in the region of non-relativistic velocities,
$v^{2}=\sum{\frac{a_{i}^{2}x_{i}^{2}}{t^{2}}}\ll 1$, the space
distributions of the thermodynamical quantities
(\ref{t}),(\ref{n}) has the Gaussian profile:
\begin{eqnarray}\label{nonrel}
\varepsilon+p_{0}  &
\simeq\frac{C_{\varepsilon}}{\prod_{i}(t+T_{i})}e^{-b_{\varepsilon}^{2}\sum{a_{i}^{2}\frac{x_{i}^{2}}{t^{2}}}},\;\label{genHubb}\\
n  & \simeq\frac{C_{n}}{\prod_{i}(t+T_{i})}e^{-b_{n}^{2}\sum{a_{i}%
^{2}\frac{x_{i}^{2}}{t^{2}}}}.\nonumber
\end{eqnarray}
The forms of the solutions (\ref{nonrel}) are similar to what was
found in Ref. \cite{ellsol} as the elliptic solutions of the
non-relativistic hydrodynamics equations. In this sense the
solution proposed could be considered as the generalization (at
vanishing pressure) of the corresponding non-relativistic
solutions allowing one to describe relativistic expansion of the
finite system into vacuum.

One can note that the case of \textit{equal} flow parameters
$T_{i}=0$ and $b_{\epsilon}=b_{n}=0$ induces formally Hubble-like
velocity profile.

It is worthily to emphasize  that the physical solutions with
non-zero constant pressure have a limited region of applicability
at least in time-like direction: if one wants to continue the
solutions to asymptotically
large times, then $(\epsilon+p_{0})_{t\rightarrow\infty}\approx\frac{C}%
{t^{3}}\rightarrow0$, and this results in non-physical asymptotical
behavior $\epsilon\rightarrow-p_{0}$, unless we set $p_{0}=0$.
Therefore, it is naturally to utilize such kind of solutions in a
region of the first order phase transition, characterized by the
constant temperature and soft EoS, $c_{s}^{2}=\partial
p/\partial\varepsilon \approx 0$, or at the final stage of the
evolution that always corresponds to the quasi-inertial flows.

\section{Conclusions}
A general analysis of the flows at soft EoS $p=const$ in the
relativistic hydrodynamics is done. A new class of analytic
solutions for 3D relativistic expansion with anisotropic flows is
found.  These solutions can describe the relativistic expansion of
the finite systems into vacuum. They can be utilized for a
description of the matter evolution in central and non-central
ultra-relativistic heavy ion collisions, especially during
deconfinement phase transition and the final stage of evolution of
hadron systems. Also, the solutions can serve as a test for
numerical codes describing 3D asymmetric flows in the relativistic
hydrodynamics.

\section*{Acknowledgments}
We are grateful to S.V. Akkelin, T. Cs\"{o}rg\H{o} and B. Luk\'{a}cs
for their interest in this work and stimulating discussions. The
research described in this publication was made possible in part by
NATO Collaborative Linkage Grant No. PST.CLG.980086, by Award No.
UKP1-2613-KV-04 of the U.S. Civilian Research $\&$ Development
Foundation for the Independent States of the Former Soviet Union
(CRDF) and Ukrainian State Fund of the Fundamental Researches,
Agreement No. F7/209-2004.


\begin{thebibliography}{99}
 \bibitem{flor}W. Broniowski, W. Florkowski, Nucl. Phys A \textbf{715}, 875c (2003);
 W. Broniowski, W. Florkowski, B. Hiller, Phys. Rev. C \textbf{68}, 034911 (2003).
 \bibitem{Busza} W. Busza, Acta Physica Polonica \textbf{B35} (2004) 2873;
 \bibitem{Hwa} R.C. Hwa, Phys.Rev. \textbf{D10} (1974) 2260; F. Cooper,
    G. Frye, E. Schonberg, Phys.Rev. \textbf{D11} (1975) 192;

 \bibitem{Bjorken}  J. D. Bjorken, Phys. Rev. {\bf D27} (1983)
   140;

 \bibitem{Chiu} C.B. Chiu, E.C.G. Sudarshan, Kuo-Hsiang Wang, Phys.Rev. \textbf{D12} (1975) 902;
 \bibitem{Csorgo} T. Cs\"{o}rg\H{o}, Heavy Ion Phys. \textbf{A21} (2004) 73-84;
\bibitem{Biro} T. S. Biro, Phys.Lett. \textbf{B474} (2000) 21-26, T. S. Biro, Phys.Lett. \textbf{B487} (2000)
    133-139;
\bibitem{AkkMunSin} S.V. Akkelin, P. Braun-Munzinger, Yu.M. Sinyukov. Nucl.Phys. \textbf{A710} (2002)
    439;

    \bibitem{Landau2} L.D. Landau, E.M. Lifshitz "Fluid
    Mechanics", Addison-Westley, Reading, Mass., 1959;
\bibitem{sin} M.I.Gorenstein,Yu.M.Sinyukov,V.I.Zhdanov: Phys.Lett. \textbf{B71} (1977)
    199, and Zh.Eksp. Theor.Fiz.- ZETF (USSR) \textbf{74} (1978), 833;

    \bibitem{ellsol}
                S.V. Akkelin, T. Cs\"{o}rg\H{o}, B. Luk\'{a}cs,
                Yu.M. Sinyukov and M. Weiner, Phys. Lett.
                \textbf{B505} (2001) 64.

    \end{thebibliography}
\end{document}